# Emerging Photoluminescence from the Dark-Exciton Phonon Replica in Monolayer WSe$_2$

Zhipeng Li[1,2#], Tianmeng Wang[1#], Chenhao Jin[3#], Zhengguang Lu[4,5], Zhen Lian[1], Yuze Meng[1,6], Mark Blei[8], Shiyuan Gao[9], Takashi Taniguchi[7], Kenji Watanabe[7], Tianhui Ren[2*], Sefaattin Tongay[8], Li Yang[9], Dmitry Smirnov[4], Ting Cao[10,11*], Su-Fei Shi[1,12*]

1. Department of Chemical and Biological Engineering, Rensselaer Polytechnic Institute, Troy, NY 12180.
2. School of Chemistry and Chemical Engineering, Shanghai Jiao Tong University, Shanghai, 200240, China.
3. Kavli Institute at Cornell for Nanoscale Science, Ithaca, NY, 14853.
4. National High Magnetic Field Lab, Tallahassee, FL, 32310.
5. Department of Physics, Florida State University, Tallahassee, Florida 32306.
6. College of Physics, Nanjing University, Nanjing, 210093, P. R. China.
7. National Institute for Materials Science, 1-1 Namiki, Tsukuba 305-0044, Japan.
8. School for Engineering of Matter, Transport and Energy, Arizona State University, Tempe, AZ 85287.
9. Department of Physics, Washington University in St. Louis, St. Louis, Missouri 63136.
10. Geballe Laboratory for Advanced Materials, Stanford University, Stanford, CA 94305.
11. Department of Materials Science and Engineering, University of Washington, Seattle, WA 98195.
12. Department of Electrical, Computer & Systems Engineering, Rensselaer Polytechnic Institute, Troy, NY 12180.
[#] These authors contributed equally to this work
[*] Corresponding author: shis2@rpi.edu, tingcao@uw.edu, thren@sjtu.edu.cn.

## Abstract

**Tungsten-based monolayer transition metal dichalcogenides host a long-lived "dark" exciton, an electron-hole pair in a spin-triplet configuration. The long lifetime and unique spin properties of the dark exciton provide exciting opportunities to explore light-matter interactions beyond electric dipole transitions. Here we demonstrate that the coupling of the dark exciton and an optically silent chiral phonon enables the intrinsic photoluminescence of the dark-exciton replica in monolayer WSe$_2$. Gate and magnetic-field dependent PL measurements unveil a circularly-polarized replica peak located below the dark exciton by 21.6 meV, equal to $E''$ phonon energy from Se vibrations. First-principles calculations show that the exciton-phonon interaction selectively couples the spin-forbidden dark exciton to the intravalley spin-allowed bright exciton, permitting the simultaneous emission of a chiral phonon and a circularly-polarized photon. Our discovery and understanding of the phonon replica reveals a chirality dictated emission channel of the phonons and photons, unveiling a new route of manipulating valley-spin.**

**Introduction**

Light-matter interactions are often dictated by optical selection rules which enable access to unique material properties such as valley-spin degrees of freedom in transition metal dichalcogenides (TMDCs)[1–10]. However, the selection rules based on electric dipole approximations also render many important processes optically "dark", prohibiting the extraction of critical material properties. To overcome this limitation, careful maneuvering of the selection rules through higher order interactions provides access to the otherwise dark states and reveal additional rich physics. To this end, two-photon luminescence spectrum probes the 2p exciton states of carbon nanotube[11] and $WS_2$[12], which provides critical information about the exciton binding energies[13–17]. Raman spectroscopy reveals the 2D vibration mode of graphene through two-phonon process, which played a significant role in identifying graphene layer number in the early stage of graphene research[18,19].

In this letter, we report the emerging circularly-polarized photoluminescence of dark-exciton phonon replica states in monolayer $WSe_2$, which arises from a second-order interaction between the spin-forbidden dark exciton and the phonon-photon degree of freedom, providing a completely new strategy to investigate the exciton-spin manifold. The chiral $E''$-phonon-assisted interaction[20–26] between the dark exciton and light leads to circularly-polarized photoluminescence (PL) from the phonon-replica state, which we unambiguously reveal and identify by the magnetic field- and gate-dependent PL spectroscopies. The energy of the $E''$ phonon mode is extracted experimentally to be 21.6 meV, which is in excellent agreement with the $E''$ phonon mode from first-principles calculations (21.8 meV, see Supplementary Note 13). We further provide a microscopic model and demonstrate the brightening of the dark exciton state through the unique role of the phonon mode, which acts as a fluctuating in-plane effective magnetic field and mixes the spin-allowed (bright) and spin-forbidden (dark) exciton states of the same valley. Our observation, therefore, opens up exciting new possibilities to probe and control the dark exciton state on its light-interaction, spin properties, and dynamics.

**Results**

**Emergence of the dark-exciton replica in monolayer $WSe_2$**

We fabricate the $BN/WSe_2/BN$ van der Waals (vdW) heterostructure through a dry pickup method which avoids exposing any vdW interface to polymer[27–29]. Two pieces of few-layer graphene were used as the contact electrode to monolayer $WSe_2$ and the transparent top gate electrode, respectively, with the top BN layer working as the dielectric. A schematic representation of the device is shown in Fig. 1a, with the optical microscope image of a device shown in the inset. With the continuous wave (CW) laser excitation centered at 1.879 eV and a low excitation power of 60 μW, low temperature (4.2 K) PL spectra of the device shown in Fig. 1b resolve distinct peaks from different excitonic complexes, including the dark exciton[30–34]. The presence of the dark exciton arises from the unique band structure of monolayer $WSe_2$, in which the conduction band minimum (CBM) and valence band maximum (VBM) are opposite in spin orientations and the lowest-energy electron-hole pairs form spin-forbidden excitons (Fig. 1d). As a result, light emission with an in-plane electric dipole is strictly forbidden, and an in-plane magnetic field[30] or the coupling to a plasmonic structure[32] is required to brighten the dark exciton. However, using an objective of large numerical aperture (N.A.), the PL of the dark exciton can still be directly observed in high-quality samples, since the radiation from a small out-of-plane dipole[33] of the dark exciton can be collected by the objective. In this case, a well-resolved dark exciton PL peak appears with a narrow linewidth. We perform valley-resolved PL spectroscopy on our devices using optical excitation with certain circular polarization ($\sigma^+$ or $\sigma^-$), and detect the PL of the same or opposite circular polarization[1,2,4,8,35–38]. Such a configuration of excitation and detection is labelled ($\sigma^\pm, \sigma^\pm$) in our work. Without applying a top gate voltage and magnetic field, the circularly polarized PL spectra in the ($\sigma^-, \sigma^-$) configuration clearly resolve the charge neutral exciton $X_0$ and two well-separated negative trions[38–41], $X_1^-$ and $X_2^-$, indicative of an initially lightly electron-doped sample. The linewidths of the two trions are 2.1 meV and 2.2 meV, respectively, much smaller than their energy splitting of ~ 7 meV[30,38–40,42–45]. It is worth noting that the linewidth of the dark exciton is as narrow as 0.9 meV, which demonstrates the quality of the spectra and is the key to our discovery of the dark-exciton replica. With the application of an out-of-plane magnetic field of 6 T, the exciton ($X_0$), trion 1 ($X_1^-$), and trion 2 ($X_2^-$) peaks all blue shift in energy and remain a single peak in the PL spectra (magenta curve in Fig. 1b), indicating that their emission is intrinsically circularly polarized at each valley. However, the dark exciton PL at 1.689 eV split into two peaks at 1.687 eV and 1.690 eV at 6 T. This splitting occurs because that the small out-of-plane electric dipole of dark excitons is expected to result in linearly polarized rather than circularly polarized light for each wavevector. Therefore, the emissions from the two valleys could both be detected in this $\sigma^-$ collection configuration, with their energy difference dictated by the valley Zeeman effect[45–48]. Remarkably, another PL peak (indicated by arrows in Fig. 1b) emerges and exhibits behavior similar to the dark excitons. This PL peak, located at 1.667 eV, splits into two peaks with unequal heights at 1.665 eV and 1.669 eV, with the application of the magnetic field of 6 T. For this reason, we label this peak as the dark-exciton replica ($X_D^R$). We have reproduced $X_D^R$ in four different BN encapsulated WSe$_2$ devices (see Supplementary Note 7 and 8). The power dependent PL intensity of the exciton, dark exciton, and dark exciton replica is shown in Fig. 1d (solid dots) at the absence of the magnetic field. The PL intensity can be fitted with a power law: $I \sim P^\alpha$, where $I$ is the PL intensity and $P$ is the excitation laser power. It is evident that the dark exciton replica and dark exciton share similar power-law exponent, different from that of the bright exciton ($X_0$). With low excitation power ($P \leq 40$ μW), the α values are 1.19 and 1.24 for the dark exciton and dark exciton replica, respectively. With higher excitation power, saturation behavior starts to occur and the α values are 0.72 and 0.77 for the dark exciton and dark exciton replica, respectively. In comparison, the PL-power scaling for the bright exciton ($X_0$) can be described with a $\alpha$ value of 1.13 throughout the whole excitation power range studied. The slight super-linear behavior of the excitation power dependence for the bright exciton ($X_0$) is consistent with previous reports[28,30].

**Magneto-PL spectra of WSe$_2$**

The connection between $X_D^R$ and $X_D$ can also be revealed through circularly-polarized magneto-PL spectra measurements taken in the $(\sigma^-,\sigma^-)$ configuration (Fig. 2). We note that the intensity oscillation as a function of the B field is a measurement artifact, which we attribute to the slight beam position shift as we increase the magnetic field. As shown in Fig. 2a, it is evident that all the peaks, except for $X_D$ and $X_D^R$, undergo a monotonic blue-shift as a function of the out-of-plane magnetic field due to the valley Zeeman effect[45–48]. On the contrary, $X_D$ and $X_D^R$ exhibit a splitting that increases linearly with the magnetic field. The emission energy in the presence of the magnetic field can be expressed as $E = E_0 \pm \frac{1}{2}g\mu_B B$, where g is the Landé g-factor of the excitonic complex of interest, $\mu_B$ the Bohr magneton. "+" and "-" correspond to the PL peak energies from the K and K' valleys, respectively. For the bright exciton, only K' valley radiation ($\sigma^-$) is allowed to be detected in the valley polarized PL spectra of the $(\sigma^-,\sigma^-)$ configuration, and hence, only the blue-shifted emission is observed. The Zeeman splitting between the two valleys, $\Delta E = E^K - E^{K'} = g\mu_B B$, is plotted in Fig. 2c (dots) where *g* factor can be obtained through a linear fitting (solid lines). The g-factor for the bright exciton, trion $X_1^-$, and trion $X_2^-$ are -3.7, -4.4, and -4.5, respectively, consistent with previous studies[28,43,44,49] and a theoretical expectation of 4 based on a non-interacting particle analysis (see Supplementary Note 3). The *g*-factor for the dark exciton, however, is -9.3, consistent with previous reports[28,31,49] and the theoretical expectation of -8 (see Supplementary Note 3). Interestingly, the dark exciton replica $X_D^R$ has a g-factor of -9.4, similar to that of the dark exciton $X_D$ but distinctly different from those of the bright exciton and trions. This particular magnetic field dependence of $X_D^R$ indicates that its spin-valley configuration is almost identical to that of the dark exciton $X_D$ (see Supplementary Note 7). It is worth noting that the relative intensity of the two branches of the dark exciton replica in the magneto-PL spectra sensitively depends on the circular polarization of bright exciton in the $(\sigma^+,\sigma^+)$ or $(\sigma^-,\sigma^-)$ measurement (Fig. 2a). The high-intensity branch switches as the helicity of circularly polarized excitation switch (see Supplementary Fig. 5e and 5f) as it closely follows the Zeeman-shifted circularly polarized bright exciton. This is in stark contrast to the two branches of the dark exciton, which are always the same in intensity regardless of the helicity of the detection. The close correlation of the high PL intensity branch of the dark exciton replica and the bright exciton strongly supports our theory of the phonon-mediated mixing of the dark exciton and bright exciton. The slight difference between the valley polarization of the dark exciton replica and the bright exciton at finite magnetic field can be potentially attributed to higher order mixing process, which is beyond the scope of this work.

**Gate-voltage dependent PL of WSe$_2$**

To further explore the origin of the replica, we investigate the PL spectra as a function of the top gate voltage for a second device, and the results are shown in Fig. 3. (The gate dependence of device 1, shown in Fig. 1 and 2, is included in Supplementary Note 8).

We employ a CW laser centered at 1.959 eV with an excitation power of 40 µW, under which the biexciton (XX) and negatively charged biexciton (XX⁻) can both be observed[28]. The $X_D^R$ peak of the second device is located at 1.676 eV, slightly higher in energy than the $X_D^R$ peak (1.667 eV) in the first device. Despite the small peak-energy shift, which possibly arises from the residual strain, the splitting between the $X_D^R$ and $X_D$ in the second device remains almost the same as that in the first device, ~ 21.3 meV. (The $X_D^R - X_D$ energy splitting value is included in the Supplementary Note 5 for all the four devices that we have measured). From Fig. 3a and 3b, it is obvious that the spectrum weight of all the resolved excitonic complexes depends sensitively on the top gate voltage that effectively controls the density and type of charge carriers in the monolayer WSe₂. While the $X^+$ occurs when the monolayer WSe₂ is hole-doped, $X_1^-$, $X_2^-$, and XX⁻ emerge when the WSe₂ is electron-doped, and XX only exists in the charge-neutral region[28,43,44,49]. It can be seen in Fig. 3a that the regions where $X_D$ and $X_D^R$ exist overlap significantly. For a quantitative understanding, we plot the integrated PL intensity as a function of the gate voltage for each excitonic complex in Fig. 3c. We find that the gate-voltage dependent integrated PL intensity of the dark exciton replica exactly mimics that of the dark exciton, both reaching the maximum near the charge-neutral region and decreasing rapidly with either electron-doping or hole-doping (gate voltage > 0 V or < -2 V, as indicated by the onset of significant PL intensity from negative trions or positive trion). The charge neutral region also strongly correlates with the PL intensity of the dark exciton. This gate-voltage dependent measurement rules out the possibility that the dark exciton replica is a charged dark exciton.

The gate-voltage-dependence and magnetic-field-and of the PL demonstrate that $X_D$ and $X_D^R$ are both charge-neutral excitations, and they should also share similar valley-spin configurations and wavefunctions. The sample-independent energy difference (~ 21.6 meV) between $X_D$ and $X_D^R$ and their sharp PL peaks further suggest that $X_D^R$, a previously unrecognized excitation of monolayer WSe₂, arises from coupling $X_D$ to a quasiparticle at ~ 21.6 meV which cannot be a charge carrier or a plasmon. Based on these analyses, we attribute $X_D^R$ to a phonon replica state formed by the coupling between $X_D$ and a phonon. To check this assumption and identify the phonon mode involved, we first perform a first-principles calculation of the phonon dispersions of monolayer WSe₂. Our results show that doubly degenerate $E''$ phonon modes appear with vibration energy $\hbar\omega_{E''} = 21.8$ meV, consistent with previous reports[24,50,51]. This energy is in excellent agreement with our observation of the $X_D$-$X_D^R$ energy difference of 21.6 meV.

**Discussion**

**Electron-phonon coupling in WSe₂**

Despite the energy agreement, it is entirely unexpected that the phonon replica PL shows an intensity comparable to the bright exciton PL (Fig. 1b and 2b), which normally would require a strong exciton-phonon coupling strength or large phonon population. Furthermore, on closer examination of the optical spectra under magnetic fields (Fig. 1b and 2b), we find that the higher-energy replica peak is much stronger than the lower-energy one, meaning that the replica PL arising from each valley has finite circular polarization. For instance, at 5 T, we estimate the degree of circular polarization of the phonon replica to be 72% (See supplementary note 6). To understand these features, we first look into the spatial symmetry of the WSe$_2$ lattice in the presence of $E''$ phonons. In Fig. 4a (inset), we schematically plot one of the doubly degenerate $E''$ lattice vibrational modes that involve the opposite in-plane movement of the upper-plane and lower-plane Se atoms. This vibration breaks the mirror symmetry about the 2D plane and the three-fold rotation symmetry. Furthermore, the vibration modes along the two in-plane directions can construct two chiral $E''$ phonon modes[50] at the gamma point, with angular momentum of 1 and -1, respectively. As a result, the excitons in the K or K' valleys can acquire a finite angular momentum and be brightened up by coupling to one of the chiral combinations (1 for K valley or -1 for K' valley) (see Supplementary Note 12). To illustrate this effect on the electrons quantitatively, we plot the conduction bands (Fig. 4a) and their in-plane spin components ($S_x$ and $S_y$) at the K valley (Fig. 4b), in a structure with the upper-plane and lower-plane Se atoms displaced along x by 0.035 Å (i.e., zero-point motion amplitude) and -0.035 Å, respectively. At the K point, we obtain a conduction band splitting of 49 meV, 9 meV larger than that of 40 meV in the equilibrium structure. More significantly, $S_x$ of the two conduction bands at K, initially 0 in the equilibrium structure, increases to $0.27\ \hbar$ and $-0.27\ \hbar$ after the displacement. Such band splitting enhancement and spin realignment clearly demonstrate an emerging in-plane spin-orbit field arising from the lattice vibrations (details and an analysis of the valence bands are included in Supplementary Note 11). As such, the $E''$ phonons are analogous to a fluctuating in-plane effective magnetic field that induces finite coupling between the two conduction bands and between the bright and dark excitons in the same valley[30], thus greatly enhancing out-of-plane and circularly-polarized PL of $X_D^R$.

**Perturbation theory of phonon-photon emission**

We further use a frozen phonon method and a second-order perturbation theory (schematically shown in Fig. 4c) to calculate the photon emission probability of the phonon replica $X_D^R$ relative to the bright exciton. The perturbing Hamiltonian consists of two terms, $H_{ex-ph}$ and $H_{ex-l}$, arising from the exciton-phonon and exciton-photon interactions, respectively. The exciton-phonon coupling allows the dark exciton to mix with the intravalley bright exciton by emitting a chiral $E''$ phonon; exciton-photon coupling describes circularly polarized light emission of the bright exciton with an in-plane electric dipole (details in Supplementary Note 10 and 12). The overall process, i.e., simultaneously emitting a chiral $E''$ phonon with an energy $\hbar\omega_{E''}$ and a circularly polarized photon with an energy $\hbar\omega_D^R$, have a transition probability given by the following equation,

$$W(\omega_D^R) = \left|\frac{\langle\Phi_D|H_{ex-ph}|\Phi_0\rangle\langle\Phi_0|H_{ex-l}|\Phi_G\rangle}{E_D - E_0 - \hbar\omega_{E''}}\right|^2 \delta\left(E_D - \hbar\omega_{E''} - \hbar\omega_D^R\right).$$

Here, $\Phi_D$, $\Phi_0$, and $\Phi_G$ are the wavefunctions of dark exciton, bright exciton, and ground state, respectively. At the experimental temperature of 4.2 K, the thermal energy of dark excitons are very small (~ 0.4 meV). We therefore use the zero-momentum dark exciton wavefunctions for $\Phi_D$. In the equation, $|\langle\Phi_0|H_{ex-l}|\Phi_G\rangle|^2$ is the photon emission probability from the bright exciton. Therefore, the other term $\left|\frac{\langle\Phi_D|H_{ex-ph}|\Phi_0\rangle}{E_D-E_0-\hbar\omega_{E''}}\right|^2$, having a dimensionless value ~0.04, defines the ratio of photon emission probability between the replica state and the bright state (calculation details in Supplementary Note 11 and 12). This large ratio, together with the long lifetime of dark excitons[30,31,52,53] of (~ 250 ± 20 ps in our device shown in Fig. 1, see Supplementary Fig. 7c), explains the significant replica PL in our experiment. It is worth noting that the dark exciton phonon replica possesses a lifetime of ~ 230 ± 20 ps, same as the dark exciton lifetime within the experimental uncertainty, confirming the phonon replica interpretation.

The dark exciton state, owing to its distinctive symmetry properties, provides an intriguing way to directly probe the "dark" phonon mode. The exciton-phonon interaction, due to the unique symmetry of the $E''$ phonon mode, leads to the formation of the dark-exciton phonon replica which potentially inherits both long lifetime of dark exciton and valley polarization of the bright exciton. Our understanding sheds light on the manipulation of dark exciton through the new knob of lattice vibrations, which also furnishes a new route for dynamically manipulating the dark exciton through the phonon-exciton interactions.

**Data availability**

The data that support the findings of this study are available from the authors on reasonable request, see author contributions for specific data sets.

**Acknowledgment**

We thank Prof. Feng Wang and Prof. Ronald Hedden for helpful discussions. This work is supported by AFOSR through Grant FA9550-18-1-0312. The device fabrication was supported by Micro and Nanofabrication Clean Room (MNCR) at Rensselaer Polytechnic Institute (RPI). Z. Li acknowledges support from the Shanghai Sailing Program (19YF1425200). S. Tongay acknowledges support from NSF DMR-1552220 and DMR 1838443. C.J. acknowledges support from a Kavli Postdoctoral Fellowship. K.W. and T.T. acknowledge support from the Elemental Strategy Initiative conducted by the MEXT, Japan and the CREST (JPMJCR15F3), JST. T.C. acknowledges support from a GLAM postdoctoral fellowship provided by the Stanford University and a start-up fund provided by the University of Washington. S.-F. Shi and Z. Lian also acknowledge support from the NY State Empire State Development's Division of Science, Technology, and Innovation (NYSTAR) through Focus Center-NY–RPI Contract C150117.

**Contributions**

S.-F. Shi conceived the experiment. Z. Li, Y. Meng, and Z. Lian fabricated the devices. Z. Li, T. Wang and Z. Lu performed the measurements. S.-F. Shi, Z. Li, T. Wang, C. Jin, T. Cao and Z. Lian analyzed the data. C. Jin and T. Cao performed calculations of the exciton-phonon coupling and developed a theoretical framework that interprets the replica photoluminescence. L. Yang and S. Gao performed DFPT calculation of the phonon modes. T. Taniguchi and K. Watanabe grew the BN crystals. S. Tongay and M. Blei grew defect-free $WSe_2$ crystals using a modified flux growth method. S.-F. Shi supervised the project. D. Smirnov supervised the magneto-PL measurements. S.-F. Shi and T. Cao wrote the manuscript with the input from all the other co-authors. All authors discussed the results and contributed to the manuscript.

**Competing financial interests**: The authors declare no competing interests.

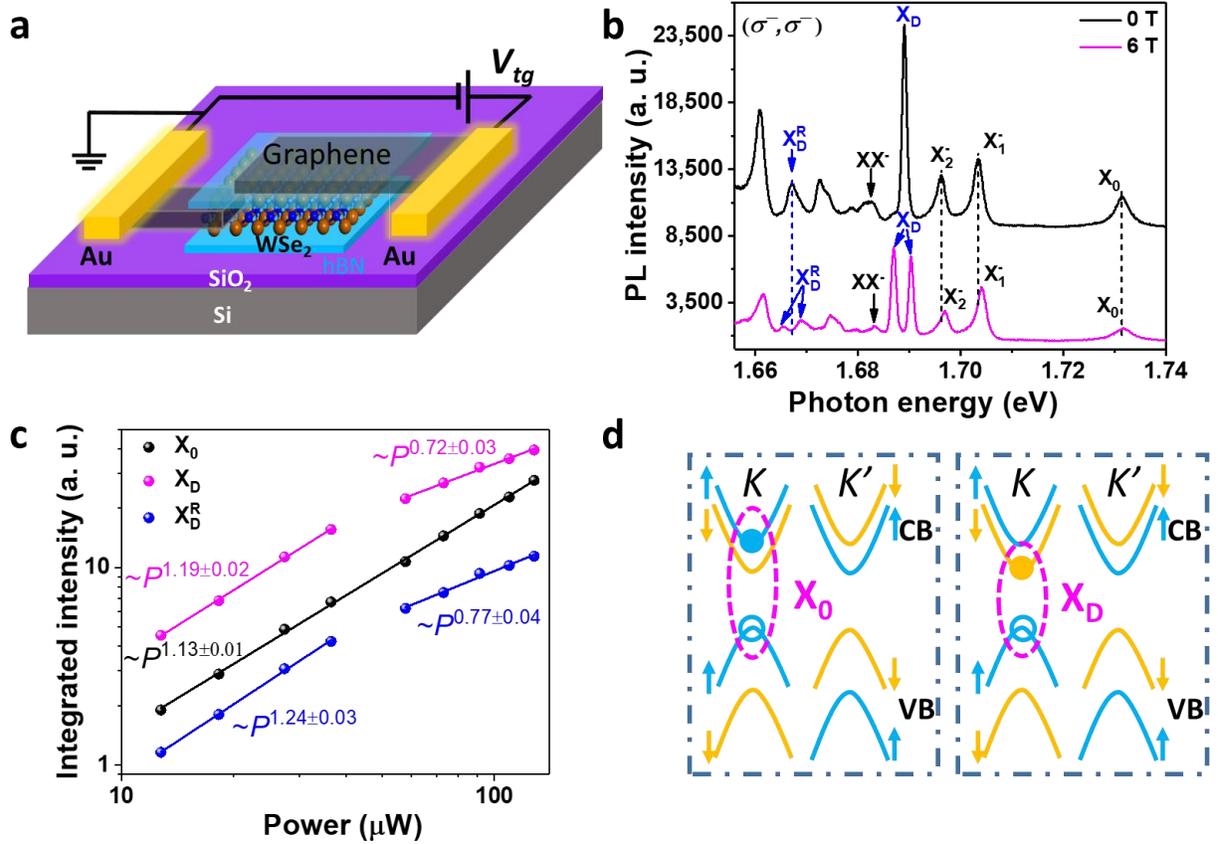

**Figure 1. Dark exciton PL splitting in magnetic field.** (a) Schematic representation of BN encapsulated monolayer $WSe_2$ with graphene contact and top gate electrodes. (b) PL spectra of the device in (a) at 4.2 K without the application of the top gate voltage, with no magnetic field (black) and with 6 T out-of-plane magnetic field (magenta) applied. (c) Integrated PL intensity of $WSe_2$ as a function of the excitation power for the $X_0$, $X_D$ and $X_D^R$ PL peaks. (d) Schematic configurations of exciton and dark-exciton states with the solid and empty dots representing the electron and hole. Blue and orange colors stand for spin-up and spin-down bands, respectively.

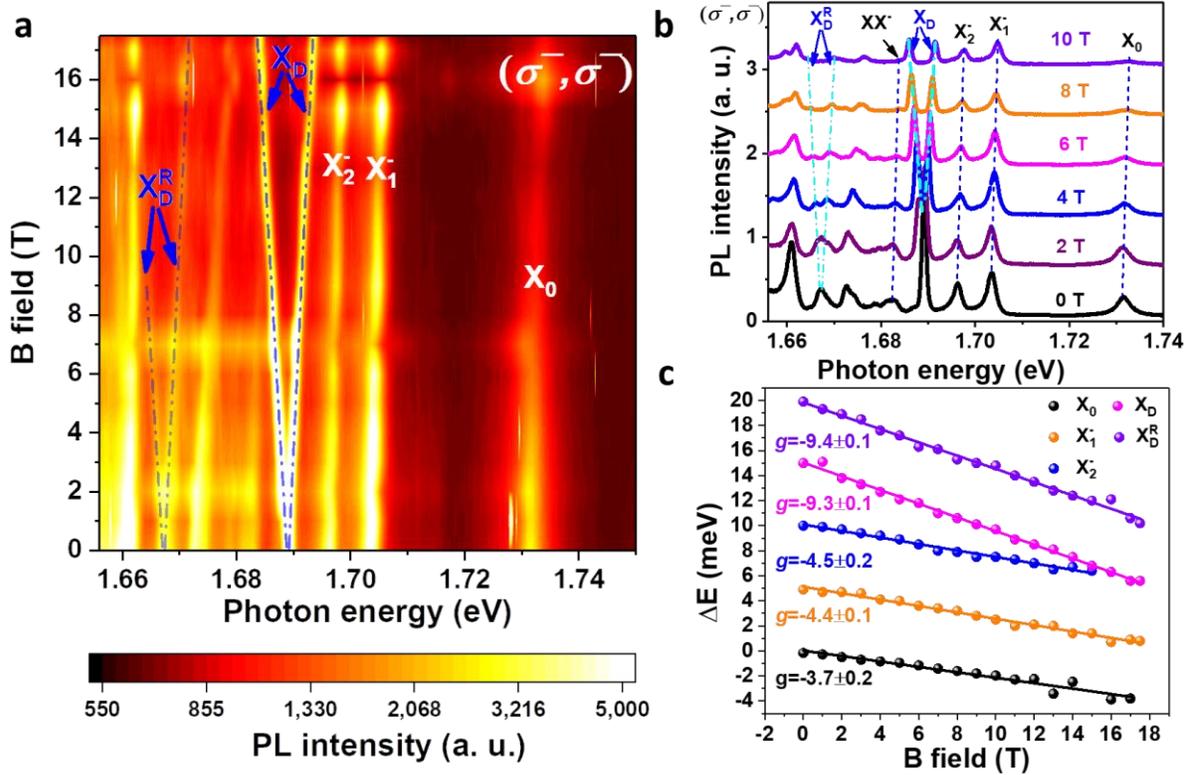

**Figure 2. Valley-resolved magneto-PL spectra of the dark-exciton and its replica.** (a) Valley-resolved PL spectra at 4.2 K as a function of the emission photon energy and the applied out-of-plane magnetic field, with the excitation of a CW laser centered at 1.879 eV and excitation power of 60 µW. The color represents the PL intensity. The dark exciton and its replica exhibit distinctively different magnetic field dependence compared to bright excitonic complexes. (b) Line traces of the PL spectra as a function of the applied magnetic field. The dashed lines are the guide for the eye. (c) g-factor for different excitonic complexes obtained from the Zeeman splitting between the $E^K$ and $E^{K'}$, obtained from (a) and Supplementary Fig. 5c. The data sets are offseted 5 meV intentionally in y-axis for clarity.

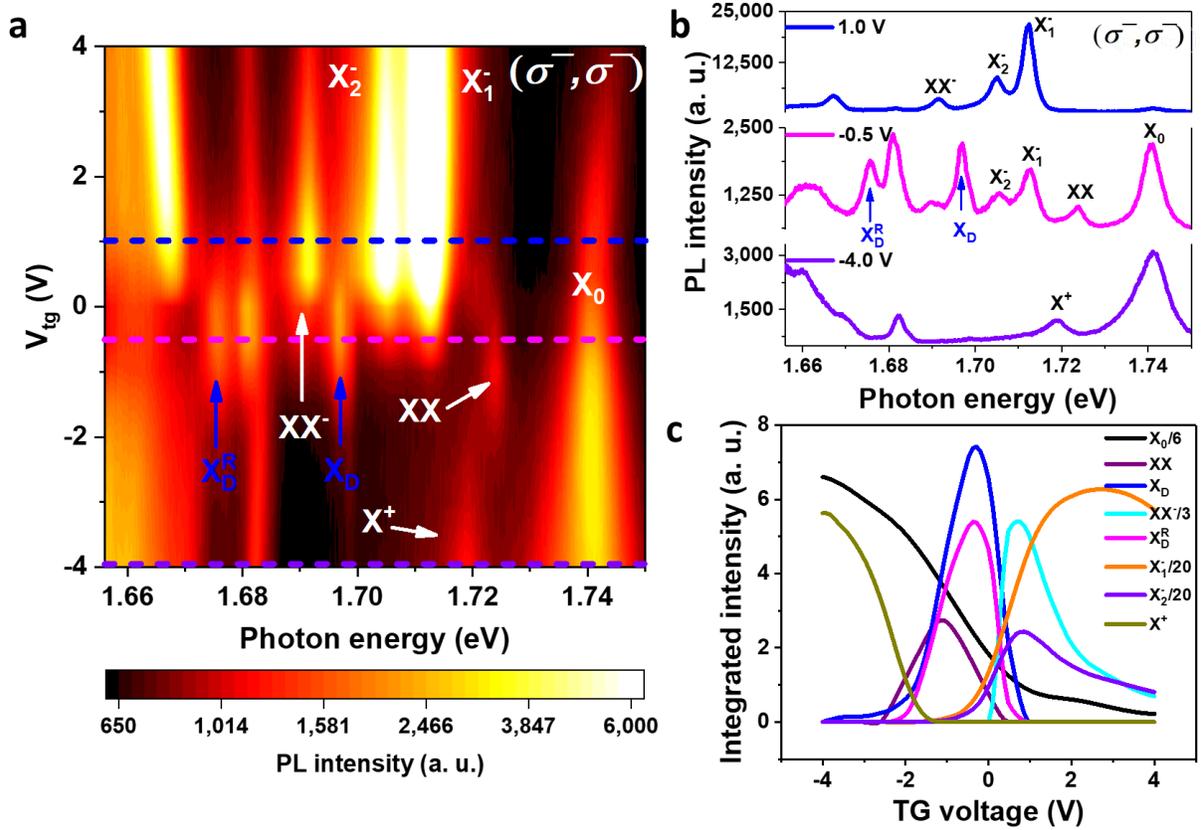

**Figure 3 Gate dependent PL intensity of the dark-exciton and its replica.** (a) PL spectra at 4.2 K as a function of the top gate voltage for a second BN encapsulated monolayer WSe$_2$ device. The color represents the PL intensity. The excitation is a CW laser centered at 1.959 eV with an excitation power of 40 µW, under which the biexciton (XX) and the charged biexciton (XX$^-$) are also visible. The gate dependence of the dark exciton replica $X_D^R$ is similar to that of the dark exciton. (b) The line traces from (a) for the gate voltage of 1.0 V (blue), -1.0 V (magenta) and -4.0 V (purple). (c) Integrated PL intensity for different exciton complexes as a function of the gate voltage. The non-zero PL intensity regions for the dark exciton and its replica are almost identical, from ~ -2.1 V to ~ 0.9 V of the top gate voltage.

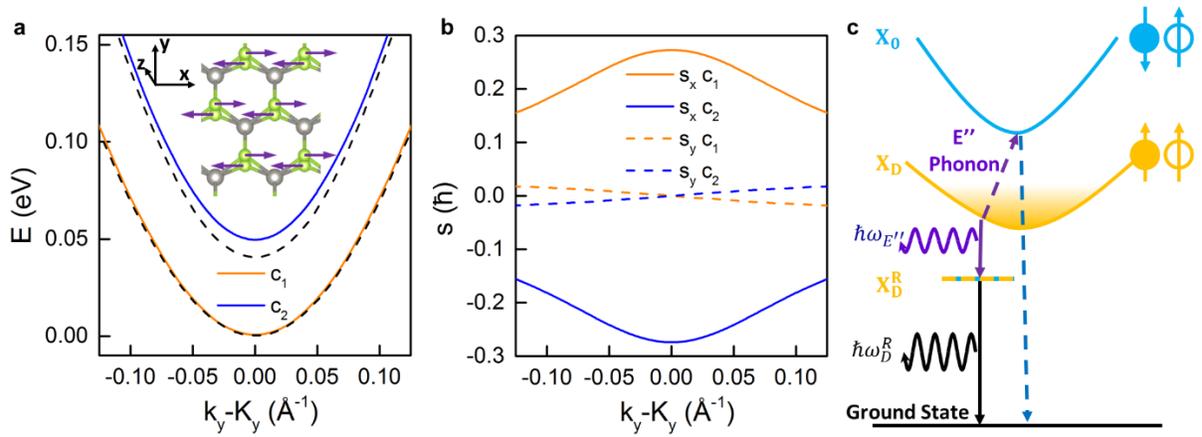

**Figure 4. Phonon coupling and recombination pathway of the K-valley dark-exciton.** (a) The conduction band structure in the K valley, with the two Se atoms in one unit cell displaced by 0.035 Å and -0.035 Å, respectively, calculated using Kohn-Sham density functional theory. The conduction band bottom is set to 0 eV. The k path is taken along the y direction across the K point, i.e., $k_x = K_x$. The lower ($c_1$) and upper ($c_2$) conduction bands are colored blue and orange, respectively. The black dashed lines are the two conduction bands in the equilibrium structure. Inset: a schematic of an $E''$ phonon eigenmode. Grey and green spheres are W and Se atoms, respectively. Arrows indicate displacement. (b) Expectation values of conduction-band electron spin angular momentum in the x and y direction, $S_x$ (solid line) and $S_y$ (dashed line), of $c_1$ (orange) and $c_2$ (blue) as a function of k. (c) The bright exciton ($X_0$) band and dark exciton ($X_D$) band are denoted by the blue and yellow parabola, respectively. The solid circle and the empty circle represent the electron and the hole, respectively, while the arrows up and down indicate the spin orientation. The yellow shaded area above $X_D$ indicates a quasi-equilibrium population of dark excitons at 4.2 K. The dark-exciton phonon replica ($X_D^R$) state is labelled by a line with alternating blue and yellow color, indicating coupling between $X_0$ and $X_D$ by emitting a chiral $E''$ phonon of an energy $\hbar\omega_{E''}$ (purple wavy arrow). The photon emission by $X_D^R$ is labelled by the black arrow, having an energy $\hbar\omega_D^R$. The emission process from $X_D$, in the second-order perturbation theory, is illustrated by the purple wavy and blue dashed lines, corresponding to the emission of a chiral $E''$ phonon and a circularly polarized photon, respectively. The intermediate state is the bright exciton $X_0$.